\documentclass[prc,aps,twocolumn]{revtex4}
\usepackage{graphicx}
\usepackage{color}
\usepackage{mathtools}
\begin{document}

\title{Temperature effects on the neutron matter equation of state         
obtained from chiral effective field theory}
\author{            
Francesca Sammarruca, R. Machleidt, and R. Millerson     }                                                           
\affiliation{ Physics Department, University of Idaho, Moscow, ID 83844-0903, U.S.A. 
}
\date{\today} 
\begin{abstract}
Temperature effects on the neutron matter equation of state are investigated 
in the framework of chiral effective field theory. Latest, state-of-the-art chiral two-nucleon forces are applied from third to fifth order in the chiral expansion together with chiral three-nucleon forces, allowing for a determination of the truncation error of the theoretical predictions. The thermodynamic quantities 
considered include the chemical potential, the internal energy, the entropy, and the free energy.
In general, good order-by-order convergence of all predictions is observed. 
As to be expected, temperature effects are largest at low density. The temperature dependence of  
the chiral three-nucleon force turns out to be weak. 
\end{abstract}
\maketitle 
        
\section{Introduction} 
\label{Intro} 
                                                                     
The nuclear equation of state (EoS) at finite temperature is of fundamental importance for a number of applications 
ranging from heavy-ion physics to nuclear astrophysics, particularly in the final stages of a supernova evolution.  
Gravitational waves in a supernova explosion may be sensitive to the temperature dependence of
 the EoS~\cite{OJ07}.        

Knowledge of the finite-temperature EoS is also critical in the interpretation of experiments aimed at identifying, for instance, a liquid-gas phase transition~\cite{JMZ84,JMZ83,SYK87,HM87,Zuo06,SMN89,Baldo95,HWW98,temp1}. 

 Earlier work on the temperature dependence of the EoS included calculations by Baldo and Ferreira who used the Bloch-De~Dominicis diagrammatic expansion \cite{temp1}, Brueckner-Hartree-Fock (BHF) calculations with 
and without three-nucleon forces (3NF)~\cite{Zuo06}, and the predictions of Ref.~\cite{FM03,FM05} based on the Green's function method.                            
Investigations of the finite temperature EoS have also been performed within the relativistic Hartree
approximation or the relativistic Hartree-Fock approach~\cite{Prak97,Ser86,Ser92,Wald87,Mull95,Glend87,WW88,MD11,RRH16}.

The entropy per particle in symmetric nuclear matter has been studied in Ref.~\cite{Rios06} within the self-consistent Green's function (SCGF) approach, where both particle-particle and hole-hole scatterings are included. The SCGF framework allows direct access to the single-particle spectral function and thus to all the one-body properties of the system. 
Work by Rios {\it et al.} \cite{Rios09} addresses hot neutron matter within the same approach and performs a comparison with other models. From that comparison, they conclude that, in the SCGF method, the internal energy per particle is generally more repulsive
than in the BHF one, but the difference is at most 1.5 MeV for the highest densities considered (0.32 fm$^{-3}$) when 
using the CD-Bonn potential~\cite{cdbonn}. 
Furthermore, for the entropy the model dependence was found to be very small, indicating that this quantity is rather
insensitive to the treatment of correlations. 
Note that the SCGF calculations of Ref.~\cite{Rios09} do not include any 3NF contributions. 

More on the phenomenological side,
hot asymmetric matter and $\beta$-stable matter have been studied by Moustakidis {\it et al.}~\cite{Moust08,Moust09} using temperature and momentum dependent effective interactions. 
Temperature effects on the EoS were also investigated within the Dirac-Brueckner-Hartree-Fock (DBHF) method. An earlier such study can be found in Ref.~\cite{HM87}.  Reference~\cite{HWW98} contains a study of isospin asymmetric matter based on the DBHF approach. One of us studied temperature effects on single-particle properties in isopsin-symmetric and -asymmetric matter within the DBHF scheme~\cite{FS10}.

During the past decade, chiral effective field theory (EFT) has become established as the best path to the development of nuclear forces, being a potentially model-independent approach with a firm link to QCD through symmetries~\cite{Wein91,Wein92,ME11} while employing nucleons and pions as basic degrees of freedom. Based on a well-defined scheme to determine the few- and many-body diagrams to be retained at each order of the expansion, chiral EFT provides a foundation for systematic studies in low-energy nuclear physics.

The temperature dependence of the EoS in isospin-symmetric 
and -asymmetric matter, as well as the symmetry energy, were investigated using microscopic two- and three-body forces derived from chiral effective field 
theory~\cite{CRP13,Car13,WHKW,WHK,WHK16,CS19,Car19}.
Those works employ the older chiral nucleon-nucleon ($NN$) potentials at N$^3$LO from Ref.~\cite{EM03} and make use of the Kohn-Luttinger-Ward~\cite{KL,LW} many-body perturbation series including contributions up to second order in the free energy.

Bearing in mind near-future applications to nuclear astrophysics, in this paper we will report on a study of the energy per particle in hot neutron matter based on most recent high-quality chiral $NN$ potentials up to fifth order (N$^4$LO) order~\cite{EMN17} of the chiral expansion, with the inclusion of an appropriately parametrized chiral 3NF. 
In neutron stars and core-collapse supernovae, the realization of the nuclear liquid-gas instability is strongly affected by the presence of leptons~\cite{WHK}.
A complete study including considerations of symmetry energy and $\beta$-stable matter at finite temperature will be the subject of forth-coming work. 

To calculate the energy in infinite matter, we sum the particle-particle contribution to infinite order through the Bethe-Goldstone equation (particle-particle ladder approximation to
account for the crucial particle-particle correlations.                                    
Temperature dependence is included in essentially two ways: 1) in the occupation number, namely, replacing the zero-temperature step function with the appropriate Fermi-Dirac distribution; and 2) in the density-dependent force which effectively describes the chiral 3NF, as detailed below. 
Item 1) is justified by the careful analysis of the Bloch-De~Dominicis diagrammatic expansion in Ref.~\cite{temp1}, where it was pointed out that the dominant terms are those corresponding to the zero-temperature Bethe-Brueckner-Goldstone expansion where temperature is included in the occupation number. 

To summarize, short-range correlations in the thermodynamical properties of hot neutron matter are included through
the extension of the BHF $G$-matrix to finite temperature. Our calculations differ from others in the literature in the choice of the chiral $NN$ interactions, the many-body method, and the treatment of the chiral 3NF. We will elaborate on our choices in the course of the paper. 

Chiral predictions must be seen in the context of the chiral order-by-order expansion which allows to determine the uncertainty of the predictions by way of the truncation error. This is one information we plan to extract fom the present study. Moreover, we are interested in the effects of temperature on the 3NF, that is, if and how the 3NF contribution is changed by the presence of finite temperature. To the best of our knowledge, systematic order-by-order studies of temperature effects with chiral forces have not been done.

The paper is organized as follows: After a description of the calculations, including a brief review of the two-body force (2NF)
and the 3NF which we employ (Sec.~\ref{calc}), we proceed to show and discuss results for the chemical potential, the internal energy, the entropy, and the free energy (Sec.~\ref{res}). Our present conclusions as well as work in progress and future plans are presented in Sec.~\ref{Concl}.

\section{Description of the calculations} 
\label{calc}
     
\subsection{The two-nucleon forces}  
\label{II} 

The $NN$ potentials employed in this work span five orders in the chiral EFT expansion, from leading order (LO) to fifth order (N$^4$LO)~\cite{EMN17}. For the construction of these potentials, the same power counting scheme and regularization procedures are applied through all orders, making this set of interactions more consistent than previous ones.  Another novel and important aspect in the construction of these new potentials is the fact that the long-range part of the interaction is fixed by the $\pi N$ LECs as determined in the recent and very accurate analysis of Ref.~\cite{Hofe+}. In fact, for all practical purposes, errors in the $\pi N$ LECs are no longer an issue with regard to uncertainty quantification. Furthemore, at the fifth (and highest) order, the $NN$ data below pion production threshold are reproduced with excellent precision ($\chi ^2$/datum = 1.15).

Iteration of the potential in the Lippmann-Schwinger equation requires cutting off high-momentum components, consistent with the fact that chiral perturbation theory amounts to building a low-momentum expansion. This is accomplished through the application of a regulator function for which the non-local form is chosen:
\begin{equation}
f(p',p) = \exp[-(p'/\Lambda)^{2n} - (p/\Lambda)^{2n}] \,,
\label{reg}
\end{equation}
where $p' \equiv |{\vec p}\,'|$ and $p \equiv |\vec p \, |$ denote the final and initial nucleon momenta in the two-nucleon center-of-mass system, respectively. In the present applications, we will 
consider values for the  cutoff parameter $\Lambda \leq 500$ MeV. 
The potentials are relatively soft as confirmed by the Weinberg eigenvalue analysis of Ref.~\cite{Hop17} and in the context of the perturbative calculations of infinite matter of  Ref.~\cite{DHS17}.

\subsection{The three-nucleon forces} 
\label{III}

Three-nucleon forces make their first appearance at the third order of the chiral expansion (N$^2$LO). At this order, the 3NF consists of three contributions~\cite{Epe02}: the long-range two-pion-exchange (2PE) graph, the medium-range one-pion exchange (1PE) diagram, and a short-range contact term. We apply these 3NFs by way of the density-dependent effective two-nucleon interactions derived in Refs.~\cite{holt09,holt10}. They are expressed in terms of the well-known non-relativistic two-body nuclear force operators and can be conveniently incorporated in the usual $NN$ partial wave formalism and the particle-particle ladder approximation for computing the EoS.       
         
The effective density-dependent two-nucleon interactions consist of six one-loop topologies. Three of them are generated from the 2PE graph of the chiral 3NF and depend on the LECs $c_{1,3,4}$, which are already present in the 2PE part of the $NN$ interaction. Two one-loop diagrams are generated from the 1PE diagram, and depend on the low-energy constant $c_D$. Finally, there is the one-loop diagram that involves the 3NF contact diagram, with LEC $c_E$. However, note that, in pure neutron matter, the contributions proportional to the LECs $c_4,c_D$, and  $c_E$ vanish~\cite{holt10,HS10}. 

The complete 3NF beyond N$^2$LO is very complex and often neglected in nuclear structure studies, although progress toward the inclusion of the subleading 3NF at N$^3$LO is underway \cite{Tew13,Dri16,DHS17,Heb15a,Kais18,Kais19}. However, there is one important component of the 3NF where complete calculations up to N$^4$LO are possible: the 2PE 3NF. In Ref.~\cite{KGE12} it was shown that the 2PE 3NF has essentially the same analytical structure at N$^2$LO, N$^3$LO, and N$^4$LO. Thus, one can add the three orders of 3NF contributions and parametrize the result in terms of effective LECs. 
In this way, one can include the 2PE parts of the 3NF up to N$^3$LO and N$^4$LO in a simple way. 
Note that among all possible 3NF contributions, the 2PE 3NF has a particular historical relevance being the first 3NF to have been calculated~\cite{FM57}. The prescriptions given above allow us to incorporate this very important 3NF up to the highest orders considered in this paper.

As explained above, we apply these 3NFs in the form of density-dependent effective two-nucleon interactions as derived in Refs.~\cite{holt09,holt10}. These effective two-nucleon potentials are 
constructed from the genuine 3NFs by summing one nucleon over occupied states. For the generalization of these expressions to finite temperature, the Fermi-Dirac occupation density given below is applied to the nucleon line that is integrated over. We use the expressions given in Ref.~\cite{WHKW}.
Thus, the calculations of this paper always  include 
the effective 2NF representing the 3NF contribution (unless stated otherwise).
In the latter, we apply the correct factors, i.~e., for the energy per particle at the Hartree-Fock level,
we apply a factor $1/3$, while for the single-particle energy, the corresponding factor is $1/2$.
Beyond Hartree-Fock there are no special factors.

\subsection{Calculations in hot neutron matter}
  
A variety of many-body methods are available and have been used extensively to calculate the EoS of nucleonic matter.
They include: the coupled-cluster method~\cite{Baa13,Hag14b}, many-body perturbation
 theory~\cite{Cor13,Cor14,Hol17,Heb11,Gez13}, and different Monte Carlo~\cite{Car15,GS16} and Green's 
function methods~\cite{FM03,FM05,Rios06,Rios09}.

As mentioned earlier, we employ the nonperturbative particle-particle (pp) ladder approximation. In the traditional hole-line expansion, it represents the leading-order contribution. In Ref.~\cite{Sam15}, we concluded that a realistic estimate of the impact of using a nonperturbative approach beyond particle-particle correlations is about $\pm$1 MeV in nuclear matter around saturation density and much smaller in neutron matter, the topic of this paper.           

In the quasi-particle approximation, the transition to the temperature-dependent case is introduced by replacing the zero-temperature occupation numbers with their finite-temperature counterparts, namely the Fermi-Dirac occupation densities.                       
 The careful analysis of the Bloch-De Dominicis~\cite{Bloch} diagrammatic expansion conducted in Ref.~\cite{temp1} concludes that the dominant terms are the ones corresponding to the zero-temperature Bethe-Brueckner-Goldstone diagrams where temperature is included in the occupation numbers. Furthermore, we include temperature effects in the density dependent force which effectively describes the 3NF, as detailed in Ref.~\cite{WHKW}.

When calculating the single-particle potential and the energy per neutron, we replace                                   
\begin{equation}
\label{step}
n(k,\rho) =\left\{
\begin{array}{l l}
1 & \quad \mbox{if $k\leq k_F$}\\
0 & \quad \mbox{otherwise } \; , 
\end{array}
\right.
\end{equation}
with the Fermi-Dirac distribution
\begin{equation}
\label{FD} 
n_{FD}(k,\rho,T) = \frac{1}{1+e^{[\epsilon(k,\rho,T)-\mu(\rho,T)]/T}} \; . 
\end{equation}
Here T is the temperature,
$\epsilon(k, \rho, T)$ the single-particle energy, function of momentum, density, and temperature, and $\mu$ the chemical potential, to 
be determined. 
The temperature-dependent angle-averaged Pauli operator is evaluated numerically. It's given by
\begin{eqnarray}
 Q(q,P;\rho,T) & = & \frac{1}{2} \int_{-1}^{+1} d(\cos\theta) \; [1 - n_{FD}(k_1,\rho,T)] \times \nonumber \\
                       &      & \times \; [1 - n_{FD}(k_2,\rho,T)] \; , 
\end{eqnarray}
for two nucleons with momenta ${\vec k}_1$ and 
${\vec k}_2$, and relative and total momentum 
given by            
${\vec q} =\frac{{\vec k_{1}} -{\vec k_2}}{2}$ and 
${\vec P} = {\vec k}_1 + {\vec k}_{2}$, respectively. 

The normalization condition 
\begin{equation}
\rho = D\frac{1}{(2 \pi)^3} \int_0^{\infty} n_{FD}(k,\rho,T) \; d^3k \; , 
\label{norm}
\end{equation}
allows to calculate the chemical potential, $\mu(\rho,T)$. $D$ is a degeneracy factor, equal to 4 for symmetric (unpolarized) nuclear matter or 2 for (unpolarized) neutron matter, our case.                      

The entropy per particle is easily obtained from the occupation probability:
\begin{eqnarray} 
\frac{S(\rho, T)}{A} & = & -\frac{1}{\rho}\frac{D}{(2\pi)^3} \int \{n_{FD}(k,\rho,T)  \;                 
\ln n_{FD}(k,\rho,T) \; +   
                            \nonumber \\           
                            &       &   + \; [1 - n_{FD}(k,\rho,T)] \times \nonumber \\
                            &       &   \times \; \ln [1 - n_{FD}(k,\rho,T)]\} \; d^3k  \; .  
\end{eqnarray}

The single-particle potential is derived from the $G$-matrix as 
\begin{eqnarray}
U(\vec{k}, \rho, T) & = &  \frac{1}{(2 \pi)^3} \int n_{FD}(k',\rho,T) \times \nonumber \\
                             &     &    \times \;  G[\vec{q}(\vec{k},\vec{k}'),                   
\vec{P}(\vec{k},\vec{k}');\rho,T] \; d^3k' \; ,
\end{eqnarray}
where the $G$-matrix is calculated by resumming the partial-wave solutions of the Bethe-Goldstone
equation for $G$ (up to total
angular momentum $j=15$) applying the
continuous choice for the single-particle potential.
The energy per particle (also known as the internal energy) is then obtained averaging the single-particle potential and kinetic energies including the 
finite-temperature occupation probability:
\begin{eqnarray}
\frac{E}{A}(\rho, T) & = & \frac{D}{\rho}  \frac{1}{(2 \pi)^3} \int n_{FD}(k,\rho,T) \; K(k) \; d^3k +
                            \nonumber \\    
                             &      &  + \; \frac{1}{2\rho}  \frac{1}{(2 \pi)^3} \int n_{FD}(k,\rho,T) \; U(k,\rho,T) \; d^3k \; ,   
                             \nonumber \\
\end{eqnarray}
where $K$ is the kinetic energy.


\begin{figure*}[!t] 
\centering
\hspace*{-3cm}
\includegraphics[width=7.7cm]{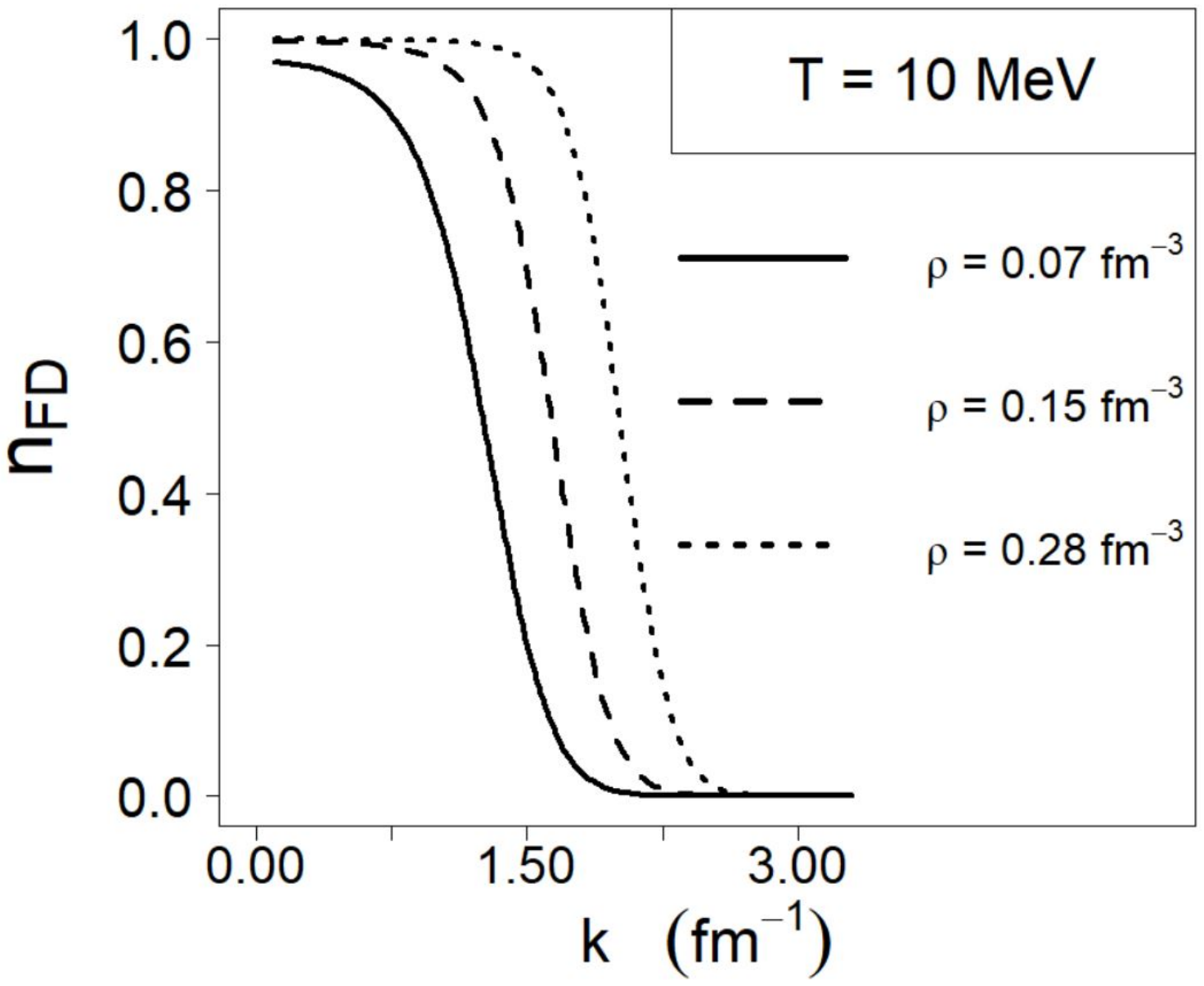}
\includegraphics[width=7.7cm]{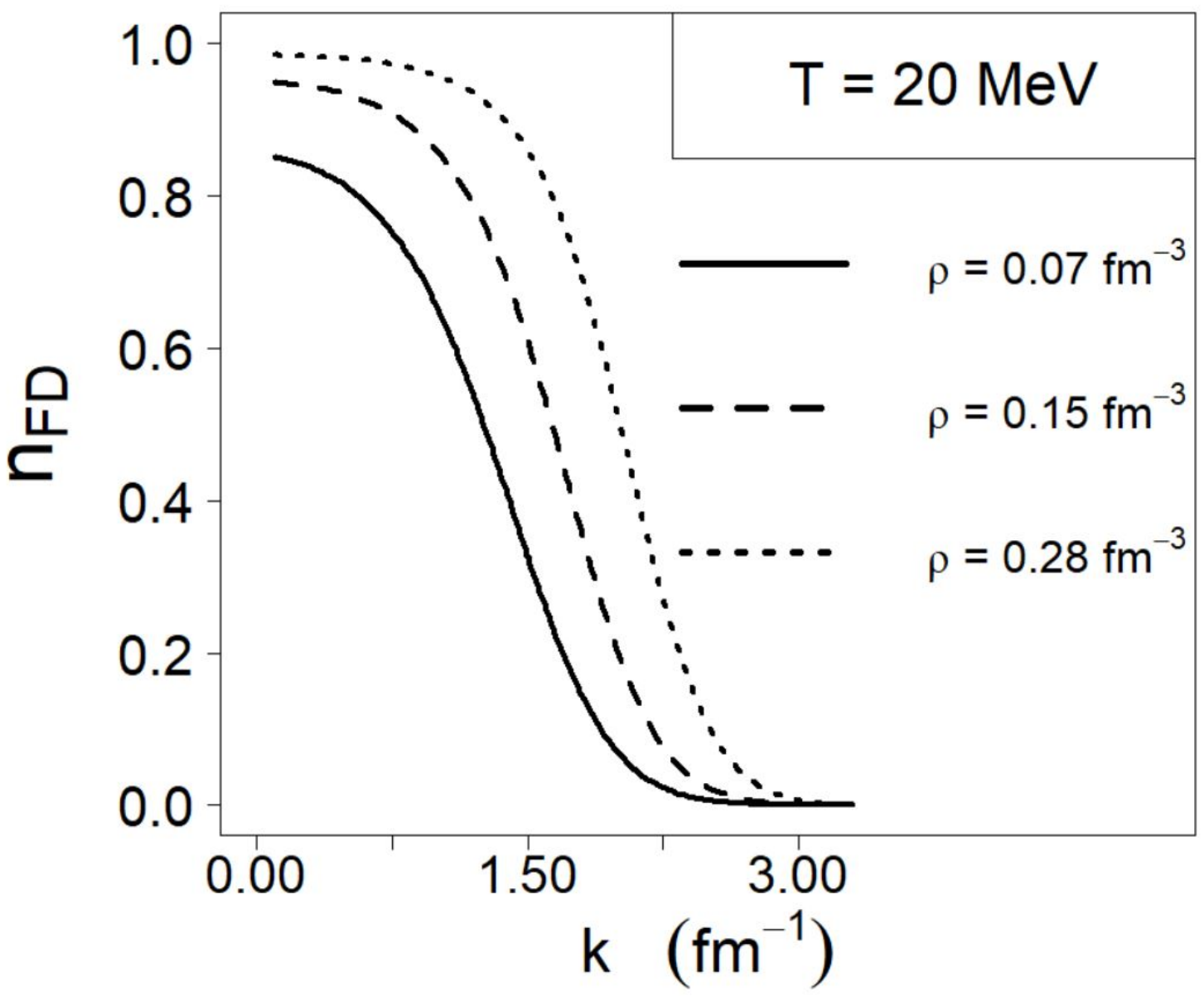}
\includegraphics[width=7.7cm]{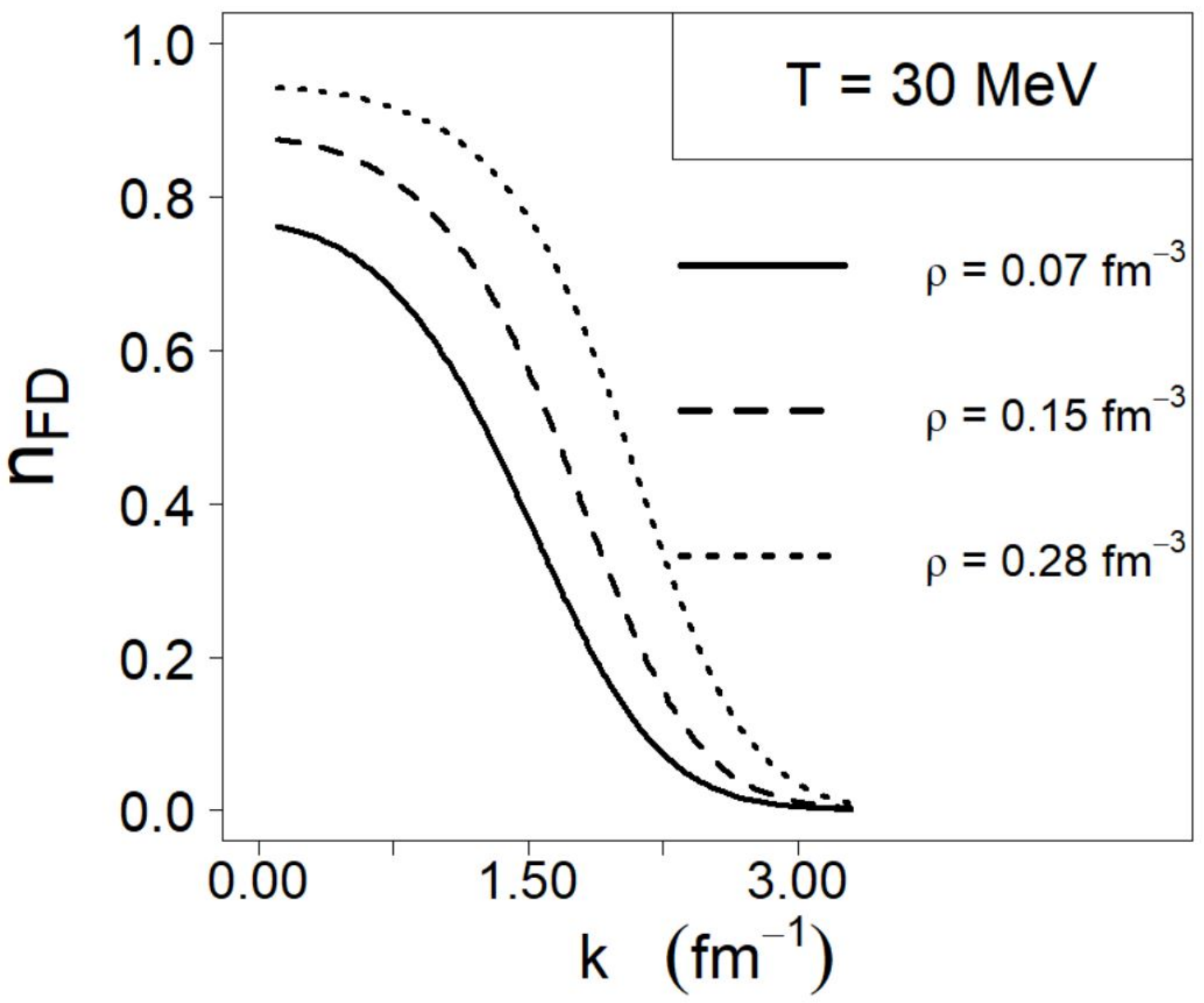}
\vspace*{-3.0cm}
\caption{Momentum distribution at T=10, 20, and 30 MeV. For each fixed-temperature case, the occupation probability
is shown for three different densities as denoted. 
The predictions are obtained with the N$^3$LO interaction using a 450 MeV cutoff.             
}                                                            
\label{FDT}
\end{figure*}

\begin{figure*}[!t] 
\centering
\includegraphics[width=8.7cm]{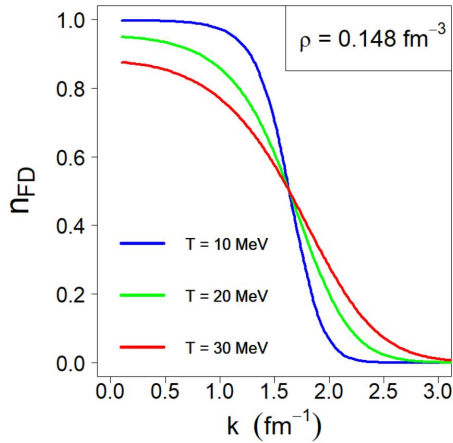}\hspace{0.01in} 
\vspace*{-3.0cm}
\caption{                                                                                     
 (Color online) The occupation probabilities at T=10, 20, and 30 MeV for a density close to normal nuclear density. The predictions are obtained with the N$^3$LO interaction using a 450 MeV cutoff.             
}                                                            
\label{FDR}
\end{figure*}

\section{Results and discussion}
\label{res} 

We will report predictions at order N$^2$LO, N$^3$LO, and N$^4$LO. We omit considerations at LO and NLO since at those low orders nuclear forces are not well described~\cite{EMN17}, and thus the corresponding predictions would not
add much to the discussion.
In fact, order N$^2$LO is the first one at which a reasonable description of nuclear forces 
becomes possible.

We begin with showing the momentum distribution, Eq.~(\ref{FD}), with varying temperature and density, see
Figs.~\ref{FDT} and \ref{FDR}. Clearly, with increasing temperature the shape departs more and more from the T=0 step 
function, more so for the low-density distributions.
Figure~\ref{FDR} displays the momentum distributions at the three temperatures and fixed density. The value of $k$ where the three curves meet is close to the Fermi momentum, which is equal to 1.638 fm$^{-1}$ for neutron matter at the shown density. Therefore, we see that decreasing the temperature (moving from the red to the blue curve) yields an increase in the occupation number below the Fermi surface while the opposite trend is seen above the Fermi level.

\begin{figure*}
    \centering
\vspace*{-0.9in}
    \begin{minipage}{0.45\textwidth}
        \centering
        \includegraphics[width=1.0\textwidth]{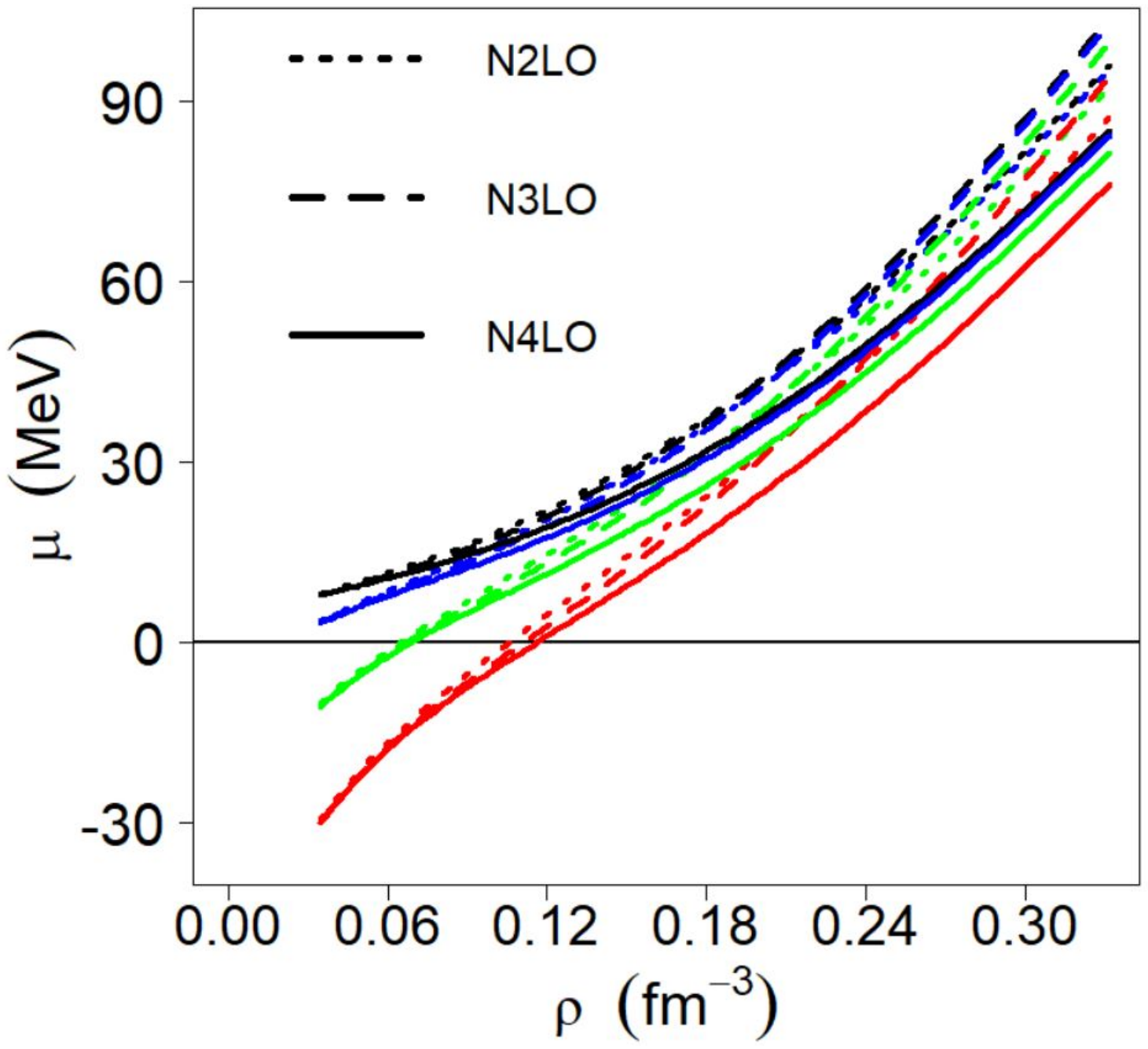}                        
    \end{minipage}\hfill
    \begin{minipage}{0.45\textwidth}
        \centering
        \includegraphics[width=0.9\textwidth]{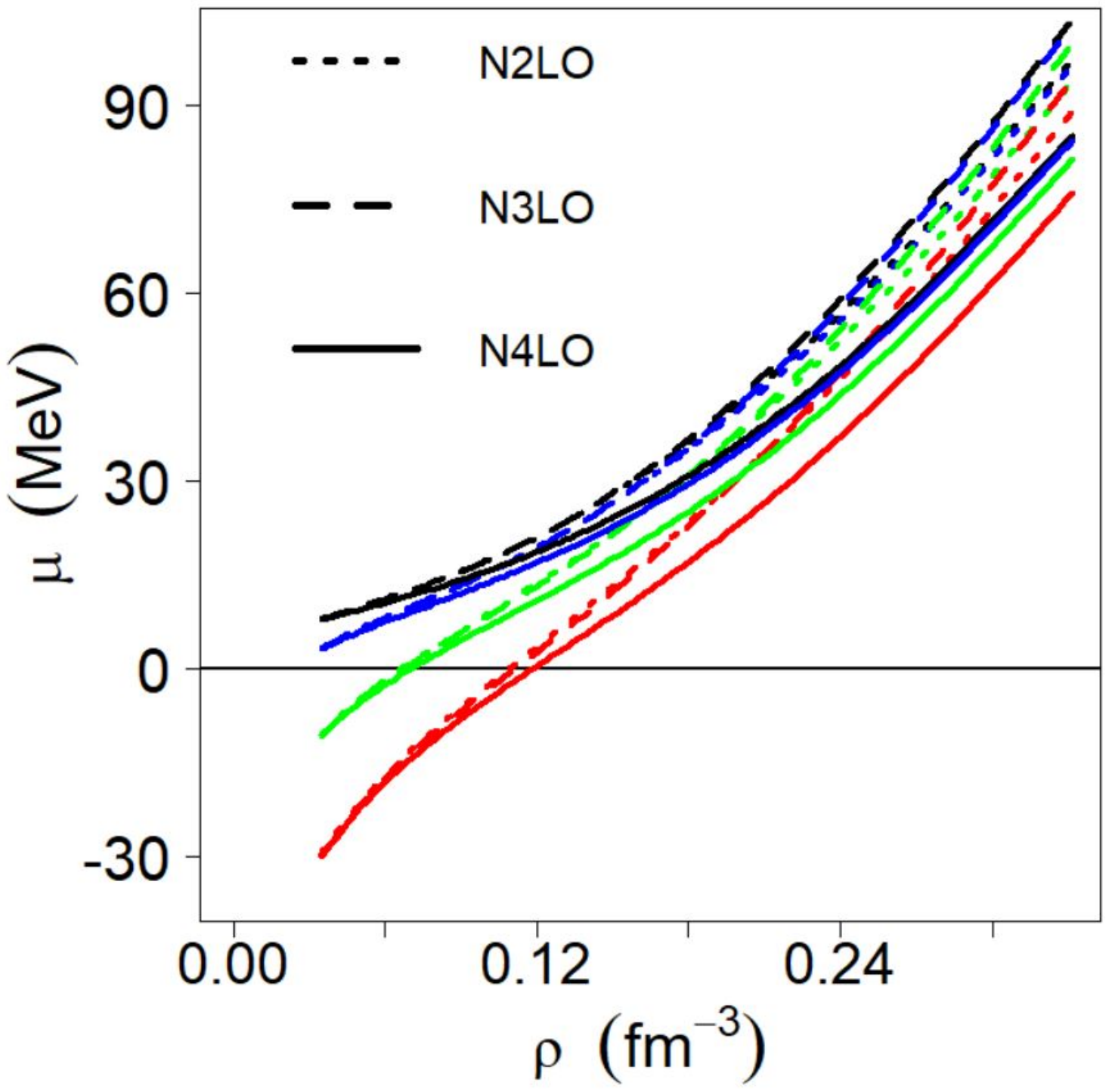}                          
    \end{minipage}
\vspace*{-1in}
 \caption{(Color online) Chemical potential in neutron matter as a function of density at the three indicated chiral orders. The black, blue, green, and red groups of curves are evaluated at T=0, 10, 20, and 30 MeV, respectively. The predictions on the left and on the right are obtained with a cutoff equal to 450 MeV and 500 MeV, respectively.
} 
 \label{nxlolam3nf9mu}
\end{figure*}

We then move on to the chemical potentials evaluated with the help of Eq.~(\ref{norm}), see 
Fig.~\ref{nxlolam3nf9mu}. The chemical potential decreases with temperature. As to be expected, temperature dependence is larger at low density, where we see excellent convergence with respect to the chiral order and well-defined temperature effects. On the other hand, at the higher densities temperature dependence becomes weaker and comparable with chiral order effects.

\begin{figure*}
    \centering
\vspace*{-0.9in}
    \begin{minipage}{0.45\textwidth}
        \centering
        \includegraphics[width=1.0\textwidth]{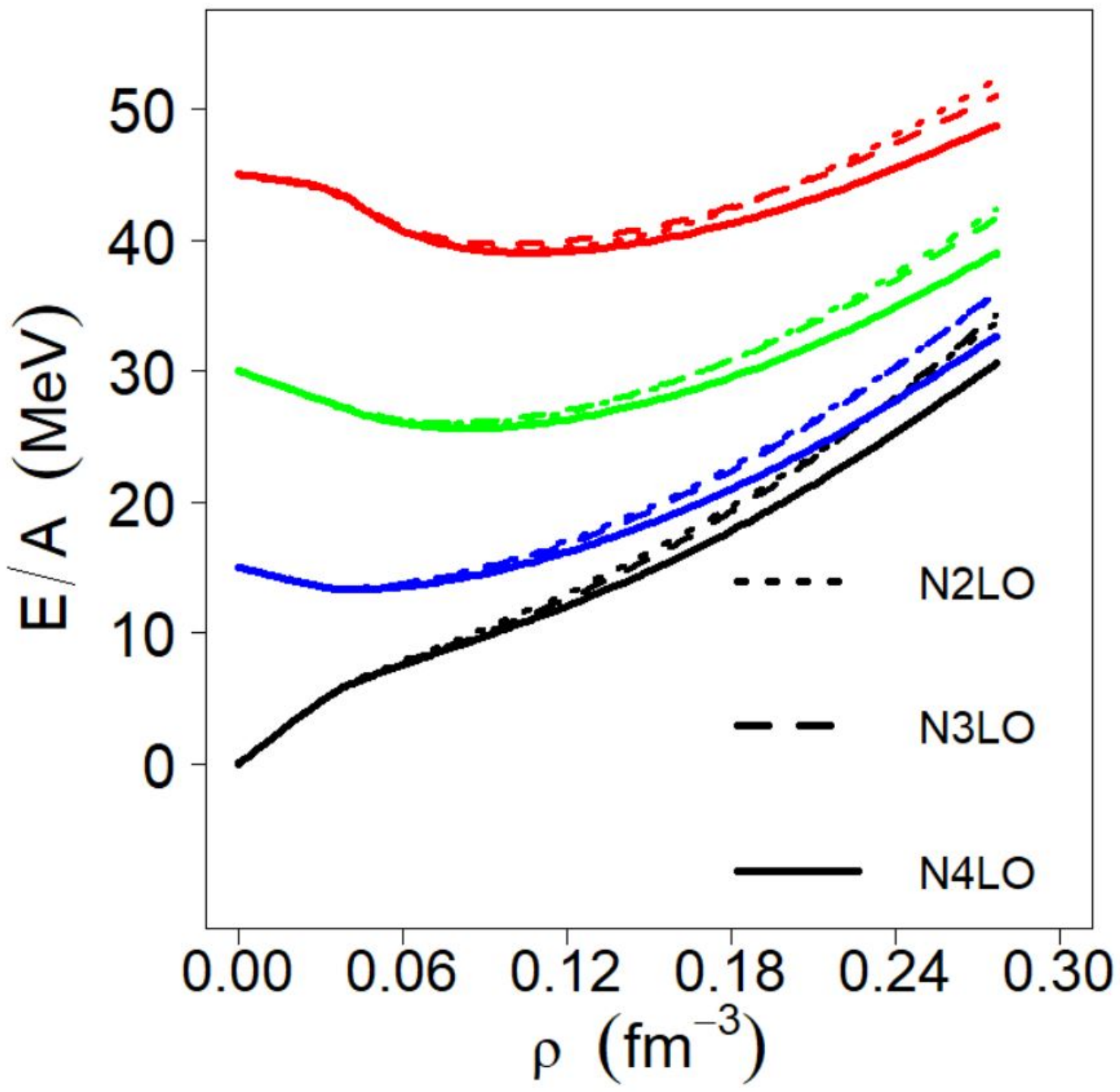}                        
    \end{minipage}\hfill
    \begin{minipage}{0.45\textwidth}
        \centering
        \includegraphics[width=0.9\textwidth]{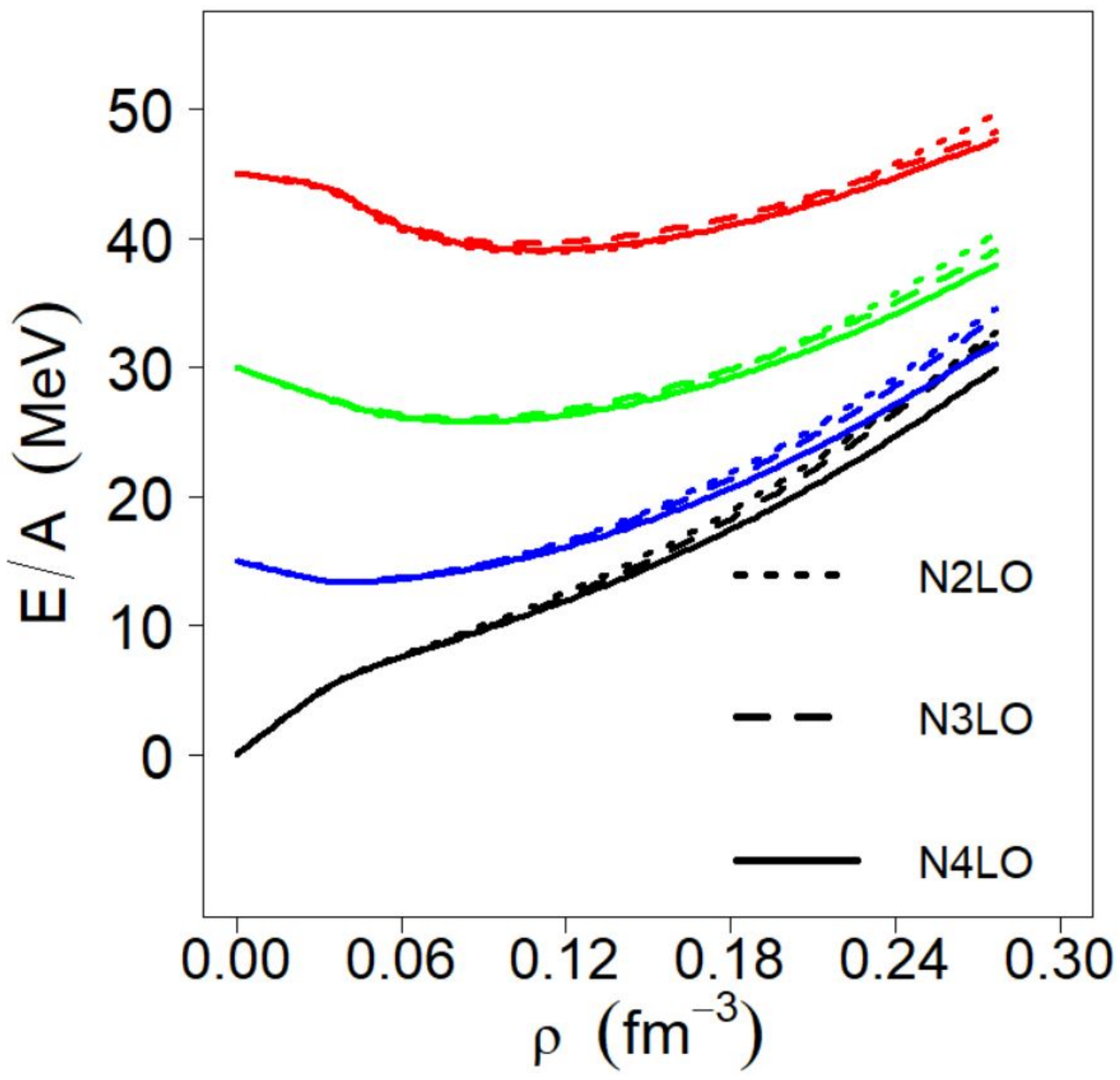}                         
    \end{minipage}
\vspace*{-1in}
\caption{(Color online) Average internal energy per particle in neutron matter matter as a function of density at the three indicated chiral orders. Color code and left and right frames as in Fig.~\ref{nxlolam3nf9mu}. 
} 
 \label{nxlolam3nf9e}
\end{figure*}

Next, we present the internal energy per particle in neutron matter, see Fig.~\ref{nxlolam3nf9e}, from T=0 to T=30 MeV. Again, temperature effects are considerably larger at the lower densities. Overall, the convergence pattern with respect to the chiral order appears satisfactory. 
At low temperature, the internal energy per particle increases monotonically as a function of density, as it should be due to the absence of liquid-gas instability in neutron matter (as opposed to symmetric nuclear matter).

A note is in place: the Brueckner-Hartree-Fock (BHF) approach we adopt for the neutron matter calculation is not suitable for very low densities. For T nearly zero, our BHF predictions are interpolated with $E/A$ = 0 at zero density, as both average kinetic and potential energies should be zero as density approaches zero.
Consistent with the virial expansion for low-density neutron matter~\cite{HS06},
 the predictions at finite temperature are interpolated such that, as the density approaches zero,
 the energy goes towards the free kinetic energy $(3/2)T$.
Note that a description of low-density neutron matter can be obtained through this virial expansion~\cite{HS06}. The handling of low-density neutron matter, however, is not one of our concerns at this time, as we are interested in astrophysical applications where appropriate crustal EoS must be attached below densities typical of homogeneous nucleonic matter.

\begin{figure*}
    \centering
\vspace*{-0.9in}
    \begin{minipage}{0.45\textwidth}
        \centering
        \includegraphics[width=1.0\textwidth]{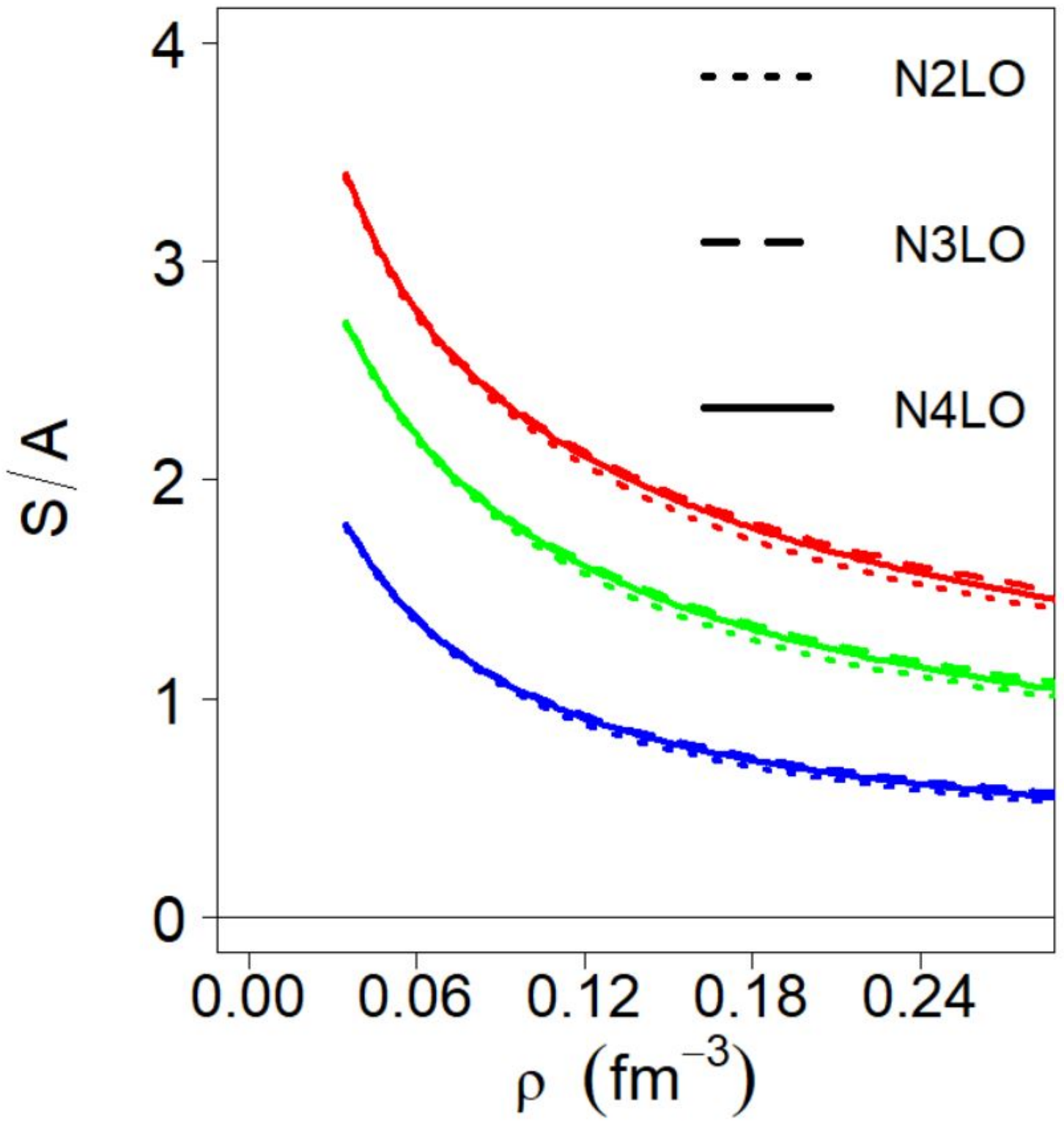}                        
    \end{minipage}\hfill
    \begin{minipage}{0.45\textwidth}
        \centering
        \includegraphics[width=0.9\textwidth]{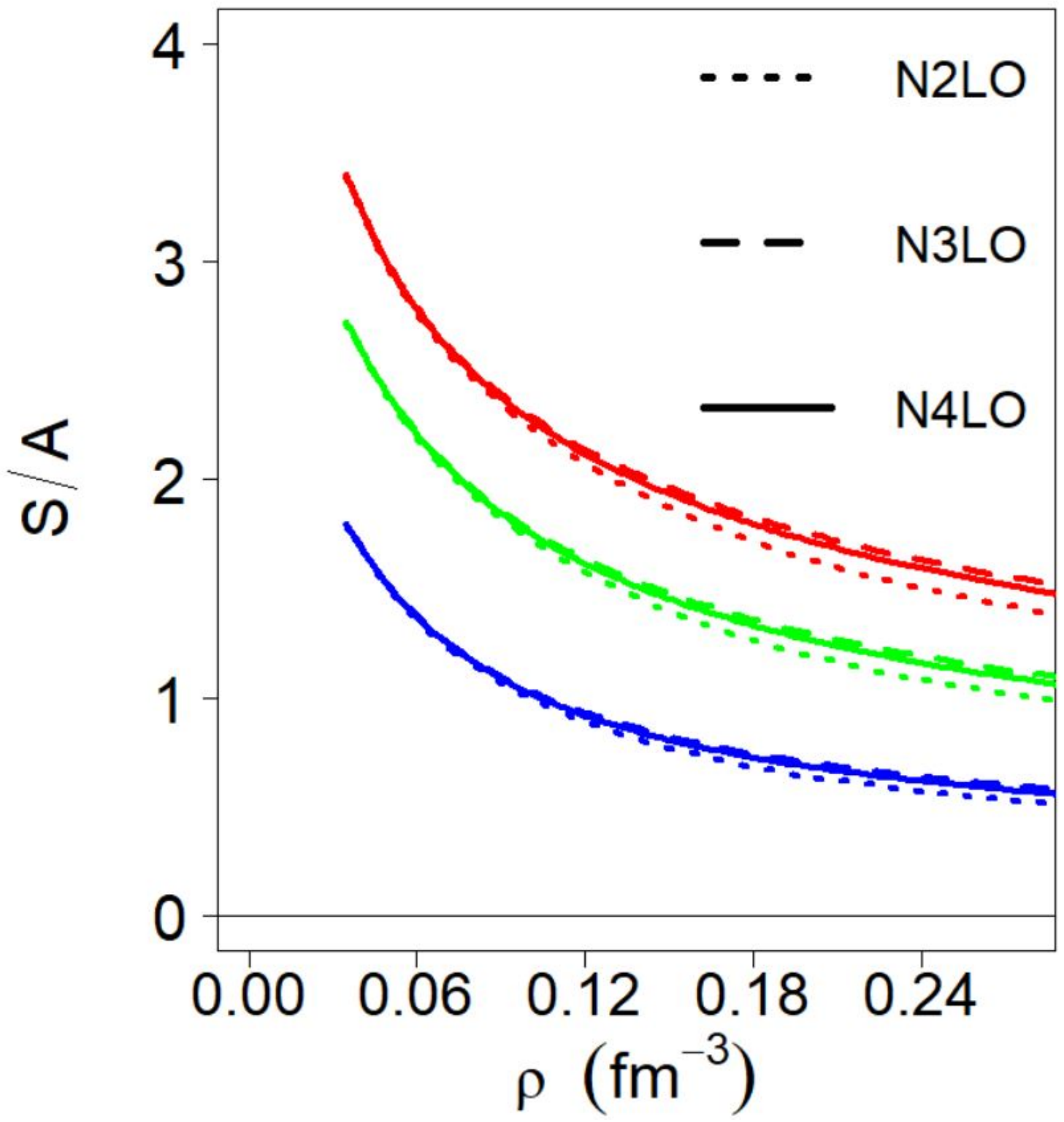}                          
    \end{minipage}
\vspace*{-1in}
\caption{(Color online) Entropy per particle in neutron matter matter as a function of density at the three indicated chiral orders. Color code and left and right frames as in Fig.~\ref{nxlolam3nf9mu}.
} 
 \label{nxlolam3nf9S}
\end{figure*}

\begin{figure*}
    \centering
\vspace*{-0.9in}
    \begin{minipage}{0.45\textwidth}
        \centering
        \includegraphics[width=1.0\textwidth]{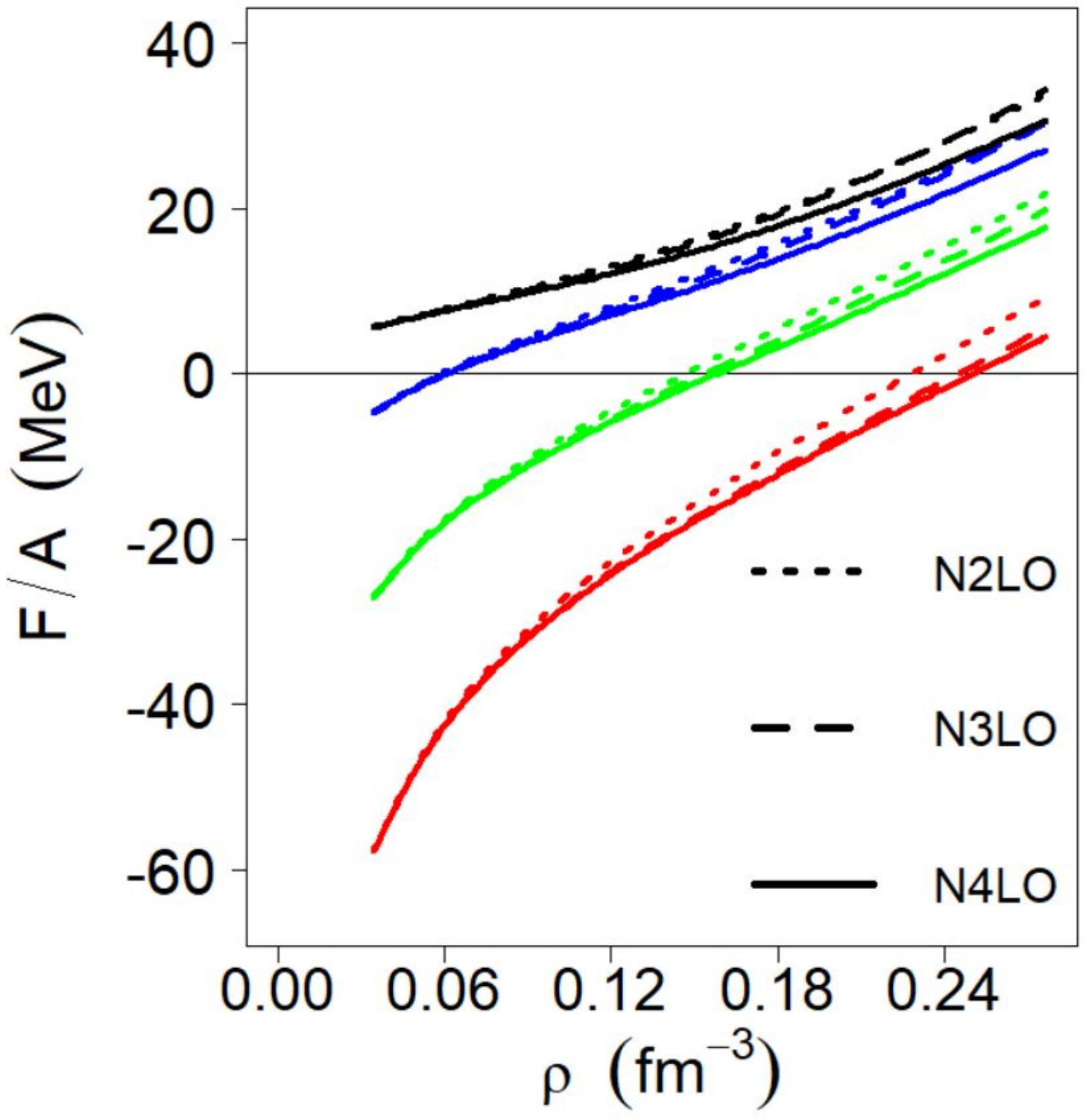}                         
    \end{minipage}\hfill
    \begin{minipage}{0.45\textwidth}
        \centering
        \includegraphics[width=0.9\textwidth]{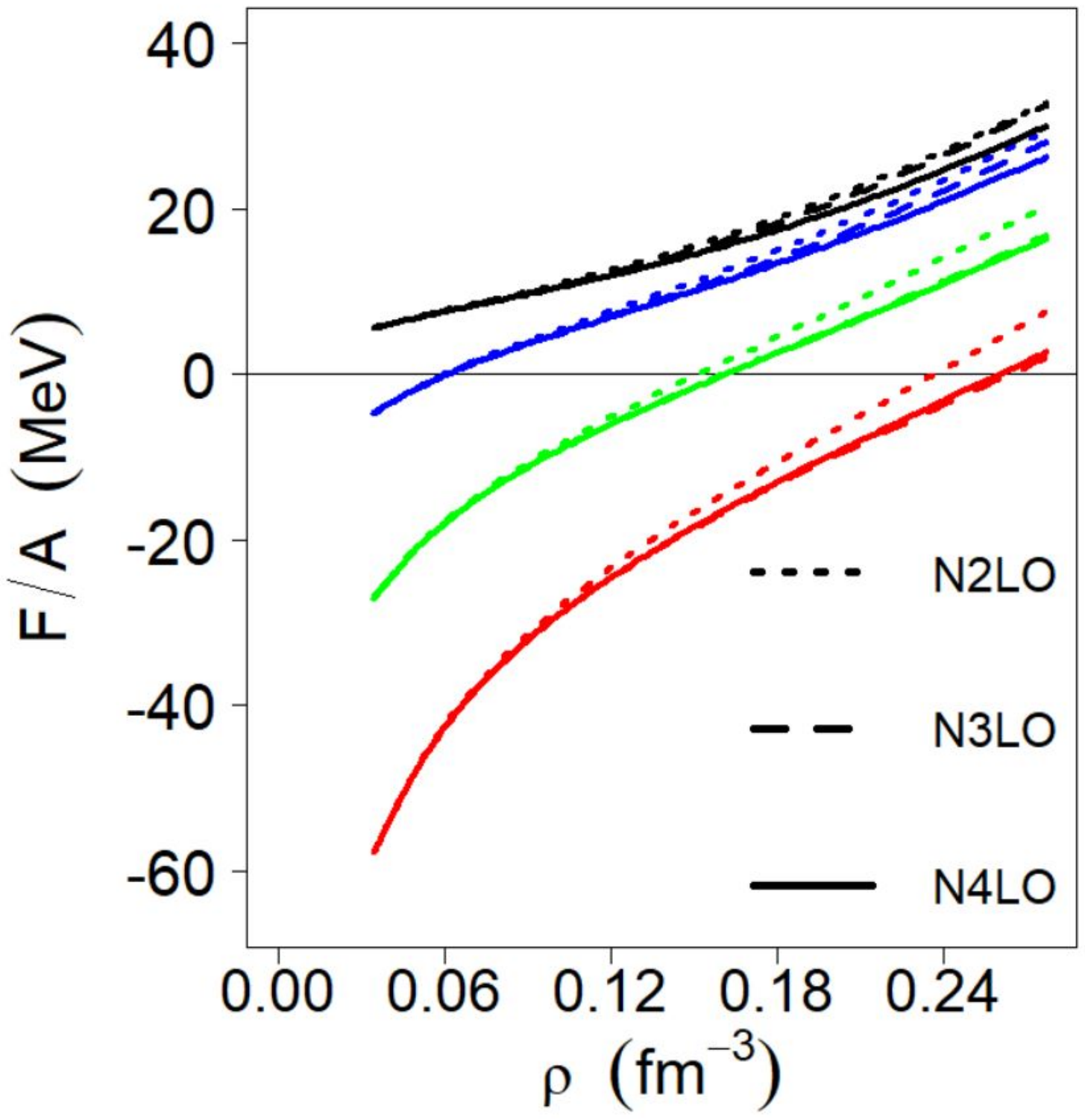}                         
    \end{minipage}
\vspace*{-1in}
\caption{(Color online) Free energy per particle in neutron matter matter as a function of density at the three indicated chiral orders. 
Color code and left and right frames as in Fig.~\ref{nxlolam3nf9mu}.
} 
 \label{nxlolam3nf9F}
\end{figure*}

The entropy per particle, $S/A$, and the free energy per particle, $F/A = E/A - TS/A$, are displayed in Fig.~\ref{nxlolam3nf9S} and Fig.~\ref{nxlolam3nf9F}, respectively, for the same interactions as in the previous figures and temperatures between 0 and 30 MeV.
The entropy, being a measure of disorder, is defined to be zero at absolute zero temperature.                             
Naturally, it increases with temperature. The temperature and density dependences of the free energy per particle show similar trends as those seen in the internal energy per particle. 

Changing the resolution scale between 450 and 500 MeV has a very small effect on all the predictions, regardless the 
temperature.

We note that the predictions for the internal energy, the free energy, and the entropy per particle appear to be in good agreement with those from Ref.~\cite{WHK}, where the Kohn-Luttinger-Ward~\cite{KL,LW} many-body perturbation series is employed. 

\begin{figure}[!t] 
\centering
\vspace*{-1in}
\includegraphics[width=8cm]{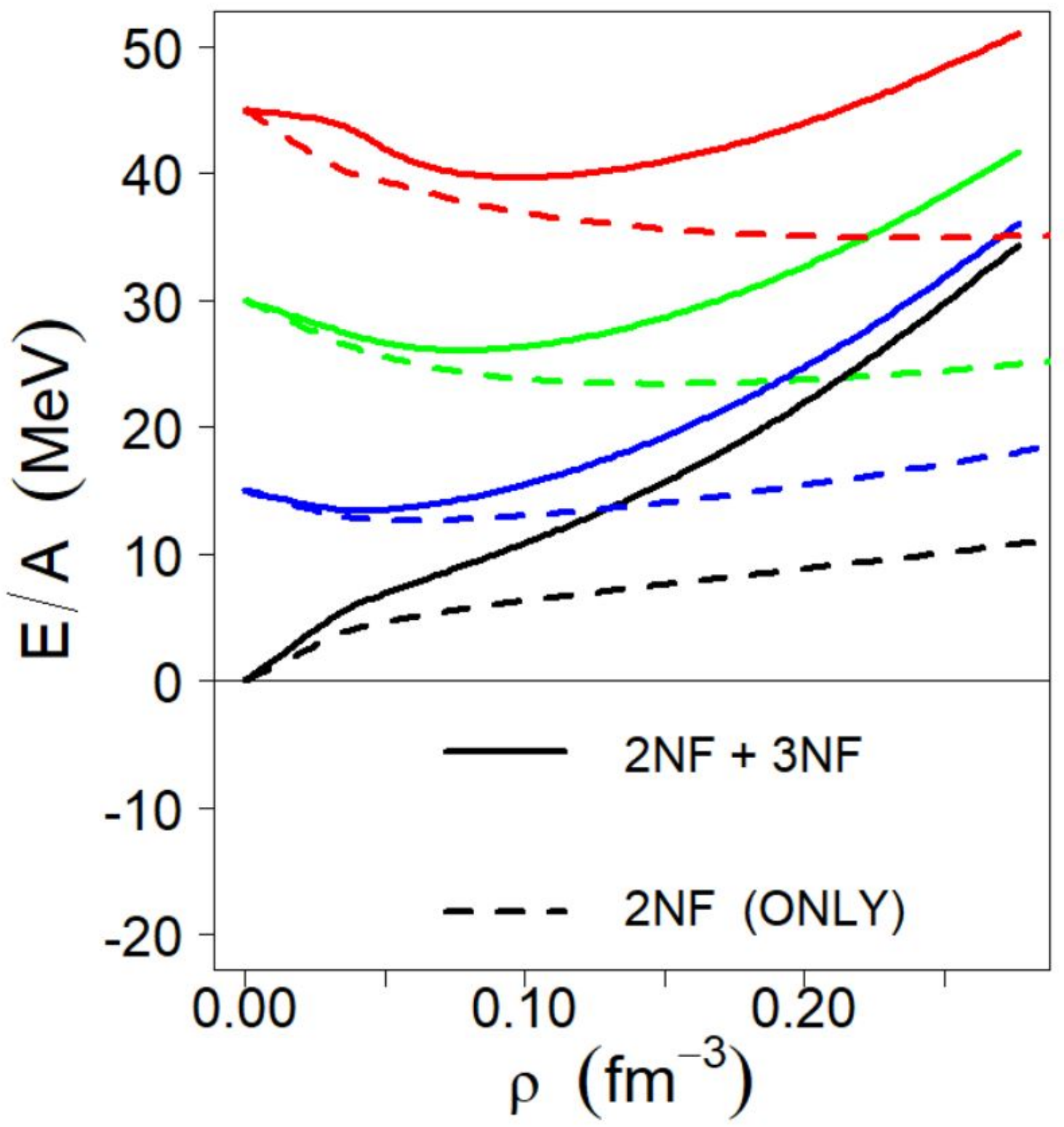}
\vspace*{-1in}
\caption{(Color online) Impact of the 3NF on the energy per particle in neutron matter matter as a function of density and varying temperature between 0 and 30 MeV. The interaction at N$^3$LO with cutoff of 450 MeV is used. Color code as in Fig.~\ref{nxlolam3nf9mu}.
} 
\label{2nf3nf}
\end{figure}

Next, we demonstrate the impact of 3NFs on the internal energy per particle with increasing temperature, Fig.~\ref{2nf3nf}.
First of all, we observe that the impact of the 3NF on the internal energy of neutron matter is large.
For the predictions based on 2NF only, the increase of the temperature tends to ``wash out", or even slightly reverse, the density dependence. This is because those predictions are not very steep functions of density to begin with, and temperature effects raise them to a larger degree at the lower densities. At high density, the differences between each curve and the one at the next higher temperature changes much more rapidly for the group including 3NFs.

\begin{figure}[!t] 
\centering
\vspace*{-1in}
\includegraphics[width=8cm]{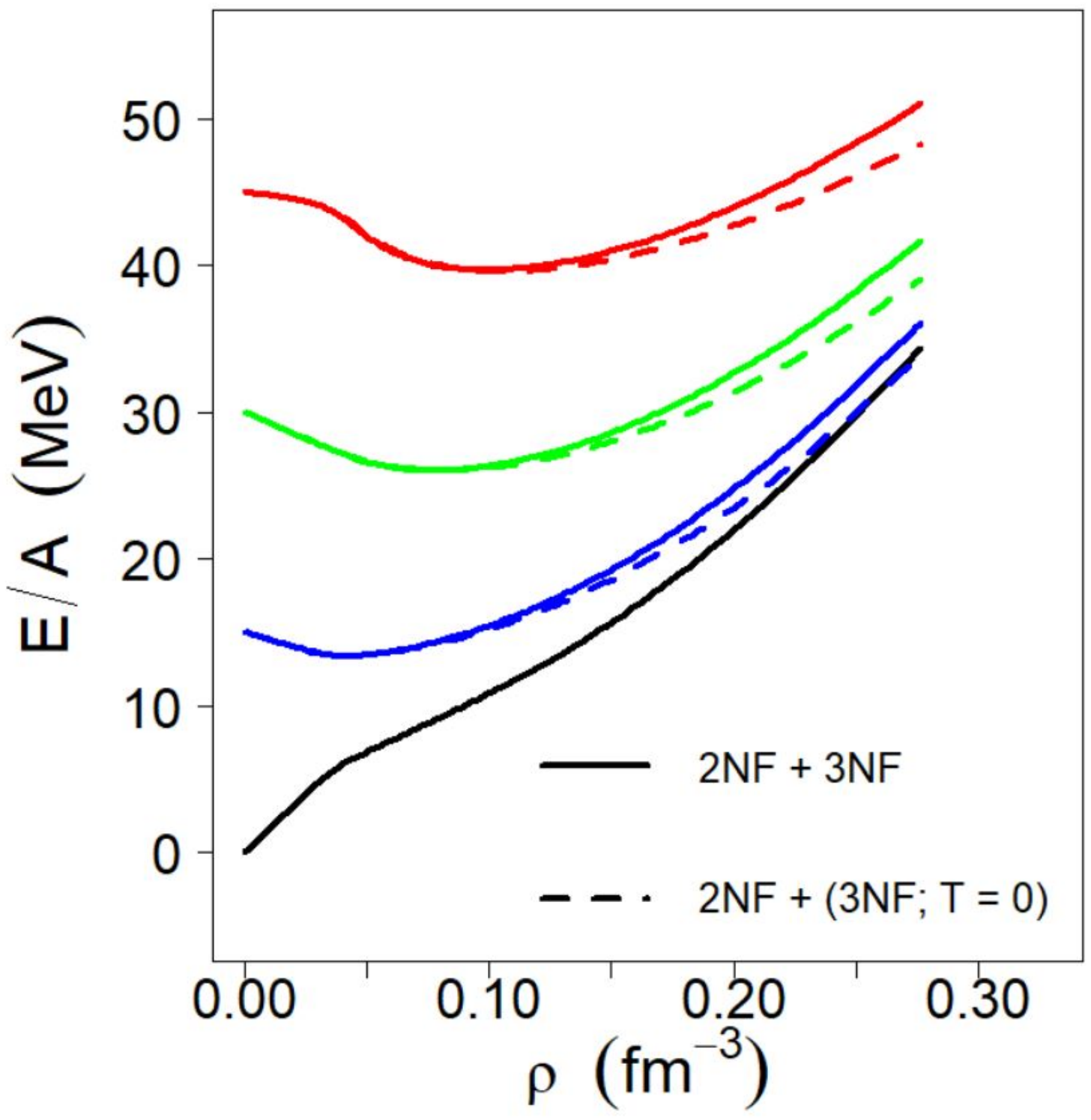}
\vspace*{-1in}
\caption{(Color online) Impact of including temperature dependence in the 3NF on the energy per particle in neutron matter matter as a function of density and varying temperature between 0 and 30 MeV. The interaction at N$^3$LO with cutoff of 450 MeV is used. Color code as in Fig.~\ref{nxlolam3nf9mu}.
} 
\label{3nft}
\end{figure}

In Fig.~\ref{3nft}, we isolate the impact of finite temperature on the 3NF only. For that purpose, we have selected 
a particular case, namely N$^3$LO with $\Lambda$ = 450 MeV, and evaluated the internal energy per particle as a 
function of density and for increasing temperature with and without temperature effects in the 3NF. 
These effects are repulsive and small, in fact 
noticible only at densities above normal density.

\begin{figure}[!t] 
\centering
\vspace*{-1in}
\includegraphics[width=8cm]{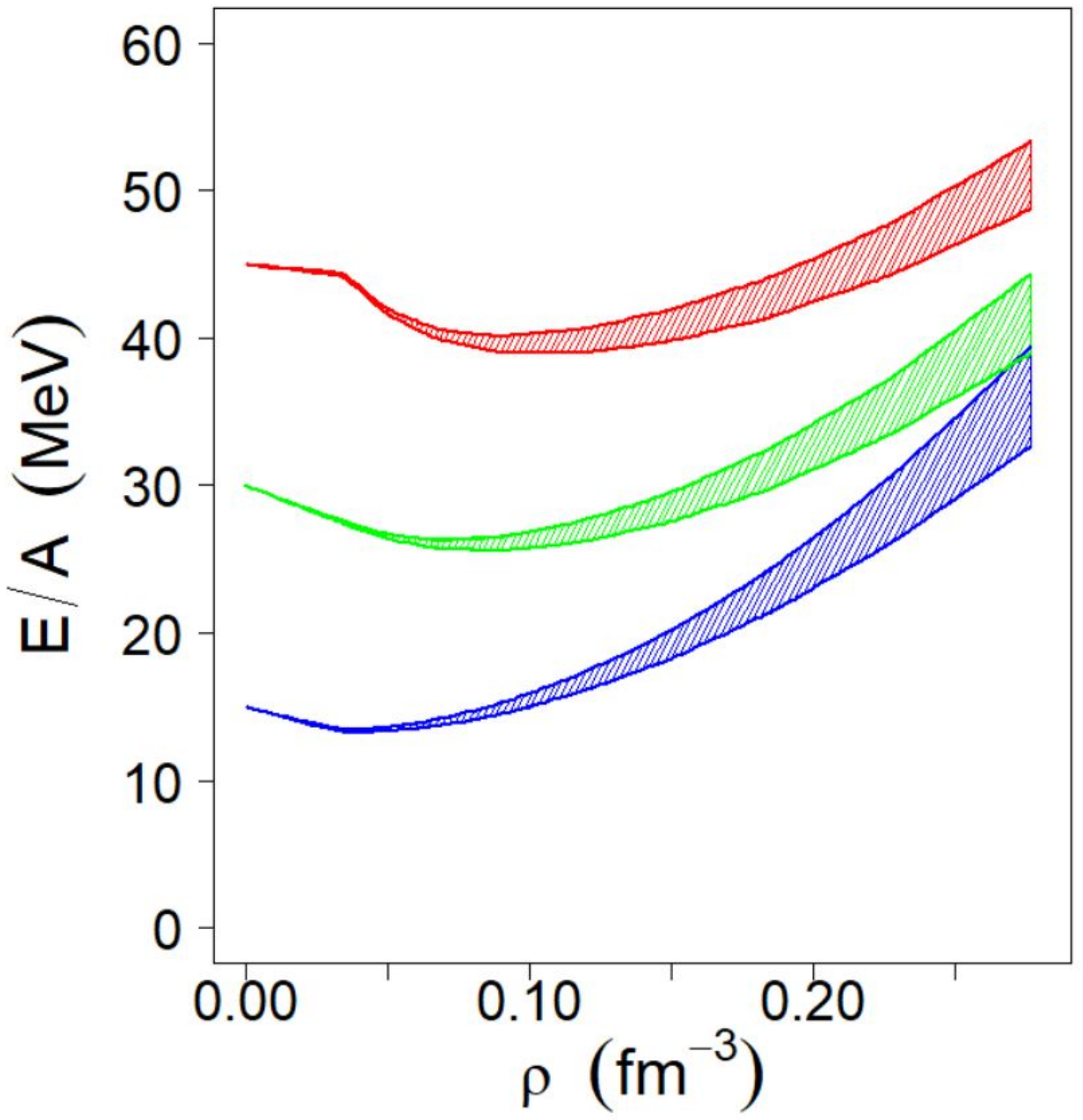}
\vspace*{-1in}
\caption{(Color online) Average internal energy per particle as a function of density.
The truncation uncertainty of the N$^3$LO predictions is calculated as in
Eq.~(\ref{eq_DeltaX}) and represented by the shaded band.                       
A cutoff of 450 MeV is used. Color code as in Fig.~\ref{nxlolam3nf9mu}.
} 
\label{bands}
\end{figure}

The purpose of this investigation is to make predictions for various thermodynamic quantities in a way which makes it possible to estimate their theoretical errors. In chiral EFT, the most important source of uncertainty is given
by the so-called truncation error which emerges from the fact that the calculation is performed only
up to a finite order of the chiral expansion. Estimating the size of the higher order contributions that are left out 
determines the truncation error at that order. If a calculation is conducted up to order $\nu$, then
 the largest ignored contribution is the one of order $(\nu + 1)$,
 which one may consider as representative for the magnitude of everything left out.
This suggests that the truncation error at order $\nu$ can reasonably be estimated to be
\begin{equation}
\Delta X_\nu = |X_\nu - X_{\nu+1}| \,,
\label{eq_DeltaX}
\end{equation}
where $X_\nu$ denotes the prediction for observable $X$ at order $\nu$.

 In view of Eq.~(\ref{eq_DeltaX}), the              
predictions at N$^4$LO will allow us to estimate the truncation uncertainty at N$^3$LO. At the fourth and fifth orders, we include the leading 3NF as explained in Sec.~\ref{III}, unless specified otherwise.
Following 
Eq.~(\ref{eq_DeltaX}), the uncertainties of our N$^3$LO predictions are given by the difference between the values 
at N$^3$LO and at N$^4$LO, which can be read off the corresponding figures. For convenience, we demonstrate in 
Fig.~\ref{bands} the uncertainty of the N$^3$LO predictions in terms of a shaded band, taking the internal 
energy as a representative observable. 

We note that more elaborate prescriptions  for estimating the truncation error
than Eq.~(\ref{eq_DeltaX}) 
 have been proposed in the literature (see, e.~g., Ref.~\cite{EKM15}), but for our estimates of the truncation errors at N$^3$LO,
this would not lead to any change, because the difference between the N$^4$LO and N$^3$LO
is larger than the one between lower orders.

\section{Conclusions and outlook}                                                                  
\label{Concl} 

In this paper, we have investigated
temperature-dependent effects on the neutron matter equation of state 
in the framework of chiral effective field theory. 
 State-of-the-art chiral two- and three-nucleon forces have been applied from third to fifth order in the chiral expansion, allowing for a thorough determination of the (truncation) error of the theoretical predictions. The thermodynamic quantities 
considered include the chemical potential, the internal energy, the entropy, and the free energy.

One of our main results is that the microscopic predictions of the temperature-dependent neutron matter EoS based upon chiral EFT show
a relatively small truncation error. Of course, only complete calculations at the respective orders allow for a 
perfect determination of the truncation error. Therefore, we note that our N$^2$LO calculations are complete at 
that order ({\it i.e.} complete 2NF plus complete 3NF). At N$^3$LO and 
N$^4$LO, we include the 2PE 3NF essentially in complete
form, but other parts of the 3NF are left out at those orders.
With the ongoing progress concerning the N$^3$LO 3NF~\cite{DHS17,Dri16,Kais18,Kais19}, it may be possible to conduct
complete calculations at N$^3$LO in the near future. In fact, preliminary evidence 
indicates that the contributions from the short-range terms~\cite{Kais18} may be negligibly small.

Furthermore, we find temperature effects on the current 3NF to be small
and noticible only at densities above normal density.

Encouraged by these trends, we plan to extend this analysis further.                     
We will investigate the thermal behavior of diverse properties, probing both low and high temperatures, such as:
\begin{enumerate}
\item Symmetry energy;
\item Symmetry energy coefficient in finite nuclei, radii, and neutron skins;
\item Location of neutron drip lines;                                             
 we wish to confront the question of drip lines for a neutron-rich nucleus                                       
immersed in a surrounding neutron sea, both at zero and finite temperature.
Such situation is expected to exist in stellar matter at subnuclear densities.
\item Stellar matter at low density and high temperatures, a scenario which is relevant for the late stage 
of a supernova collapse.
\item Liquid-gas phase transition in isospin-asymmetric matter.
\end{enumerate}

All of the items above will be analysed within the context of modern interactions of the highest quality and 
a realistic quantification of the theoretical uncertainties. We believe that both these elements are 
essential to guide experiments.

\section*{Acknowledgments}
This work was supported by 
the U.S. Department of Energy, Office of Science, Office of Basic Energy Sciences, under Award Number DE-FG02-03ER41270.

\end{document}